# Scalable Software Testing in Fast Virtual Platforms: Leveraging SystemC, QEMU and Containerization

An AI Accelerator Example

Lukas Jünger, MachineWare GmbH, Aachen, Germany (*lukas@mwa.re*)

Jan Henrik Weinstock, MachineWare GmbH, Aachen, Germany (*jan@mwa.re*)

Tim Kraus, Robert Bosch GmbH, Renningen, Germany (*tim.kraus@de.bosch.com*)

*Abstract*—The ever-increasing complexity of HW/SW systems presents a persistent challenge, particularly in safety-critical domains like automotive, where extensive testing is imperative. However, the availability of hardware often lags behind, hindering early-stage software development. To address this, Virtual Platforms (VPs) based on the SystemC TLM-2.0 standard have emerged as a pivotal solution, enabling pre-silicon execution and testing of unmodified target software. In this study, we propose an approach leveraging containerization to encapsulate VPs in order to reduce environment dependencies and enable cloud deployment for fast, parallelized test execution, as well as open-source VP technologies such as QEMU and VCML to obviate the need for seat licenses. To demonstrate the efficacy of our approach, we present an Artificial Intelligence (AI) accelerator VP case study. Through our research, we offer a robust solution to address the challenges posed by the complexity of HW/SW systems, with practical implications for accelerating HW/SW co-development.

## I. INTRODUCTION

The landscape of embedded HW/SW platforms has undergone a remarkable transformation, marked by a surge in complexity and variety. This is primarily attributed to a lower barrier of entry into custom hardware design, a trend facilitated by advancements in technology, methodologies and new tools. As a result, we witness not only a proliferation of custom hardware systems but also a corresponding increase in the sophistication of the executed software. However, this proliferation has brought about a pressing demand for flexible and expedited evaluation and verification processes.

Among the key challenges encountered in this domain is the urgent need for early software development and testing, as well as the assessment of system performance, which necessitate the availability of accessible and scalable simulation approaches. The restrictive nature of commercial simulator licenses is a significant challenge in that domain. Despite the availability of powerful simulation tools, their broad adoption is hindered by limited seats and contractual durations that fail to align with the timeframe of product development. Consequently, there exists a demand for simulation solutions that offer greater accessibility and adaptability.

In response to these challenges, this paper proposes a novel approach towards achieving flexible, accessible and scalable simulation. By addressing the limitations posed by restrictive commercial license models and emphasizing the importance of customization as well as portability, we aim to facilitate agile development in the realm of embedded HW/SW systems. To fulfill this, we propose an approach that utilizes containerization to encapsulate the entire simulation environment, coupled with an open and license-free Virtual Platform (VP) architecture constructed on the foundation of the open-source SystemC TLM productivity library VCML [2] and the widely used, open-source QEMU [3] simulator. The outcome is a high-performance VP, seamlessly integrated within a scalable framework, tailored to meet the demands of contemporary testing and verification processes.

## II. RELATED WORK

QEMU is a robust, open-source system simulator featuring Dynamic Binary Translation (DBT)-based instruction set simulation and a diverse array of peripheral models, including UARTs, CAN controllers, and Ethernet controllers. Its open-source nature has led to widespread adoption across various embedded software development and verification scenarios, such as fault injection [4], code coverage analysis [5], and security use cases like fuzz testing [6]. Despite its versatility, QEMU is hindered by inherent limitations stemming from its legacy, monolithic architecture. Key challenges include a lack of modularity and standard interfaces for integrating non-QEMU simulation models. These issues have been documented in several studies that integrate QEMU with standard hardware modeling solutions like SystemC TLM-2.0 [7-10]. These integrations serve as foundational steps, facilitating SystemC TLM-2.0 co-simulation for a subset of QEMU models.

### III. SCALABLE EXECUTION FRAMEWORK

The typical workflow for pre-silicon software development requires extensive initial effort to set up the specific work environment, consisting of software development components as well as the simulation tools and models. In practice these preparation steps are time-consuming and challenging. Tool requirements in terms of operating systems, 3rd party dependencies and component versions are extensive and often conflicting. Deviating environments between supplier and user cause ambiguities that require case-by-case debugging. All this leads to a long and hard to predict *time-to-execution (TTE)* for the VP.

To overcome this, containerization technology, such as Docker, is a powerful instrument. In this work, we are making use of the SUNRISE (*Scalable Unified RESTful Infrastructure for System Evaluation*) framework [11]. As shown in Figure 1, the infrastructure is built around a central *Runtime Manager* that interacts with the VP containers through a standardized API to configure simulation parameters, execute it and retrieve result artifacts. By encapsulating a simulation in a container image, the responsibility for creating the VP execution environment remains with the simulation provider, who has most of the VP expertise. The complete environment with all dependencies is contained in the docker image, which is either delivered ready-to run to the user or in form of the recipe, e.g. a Dockerfile, so the image can be built on the user's side.

Additionally, containerization introduces scaling capabilities, since it enables the efficient and interference-free replication of the same VP multiple times in parallel. This capability is particularly advantageous for running multiple instances simultaneously in verification scenarios.

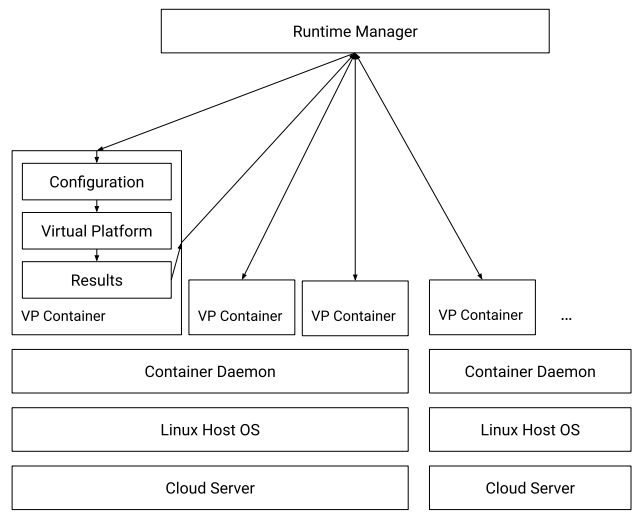

Figure 1: Execution Environment Overview

### IV. QEMU IN SYSTEMC

The central component of embedded systems is the CPU sub-system, containing the core and tightly coupled peripherals. In VPs, this part can be modeled with a variety of approaches, each different in terms of timing accuracy and simulation performance.

A well-established solution is QEMU [3], which is ideally suited to address the challenges of modern HW/SW co-design. Its open-source repository features many different core and peripheral models, making it highly adaptable. The employed technology and open-source licensing model facilitate encapsulation in a container image and therefore the scaling of execution instances as no seat licenses are required to execute the simulation.

To leverage CPU-subsystems from QEMU in VPs together with existing and future hardware models, like peripherals and other components on printed circuit board (PCB) level, a connection to Accellera SystemC is paramount. The SystemC/TLM standard serves as the industry-recognized approach for achieving this integration. *QBox* enables QEMU/SystemC integration and seamless use of all QEMU models in SystemC without compromising any performance. This allows connecting any custom SystemC modules that also comply to the standard and therefore is the key property for good expandability for future use-cases. By integrating with VCML, the effort for constructing new SystemC TLM-2.0 models can be reduced, as model building blocks such as register interfaces, configuration properties or TLM-based interconnect protocols are readily available. Furthermore, VCML offers different output options for data that is generated during simulation. Besides printing to the screen, data may be stored into files or made available through network sockets. This simplifies containerization and head-less use of VCML-based VPs in general.

## V. CASE STUDY: AI ACCELERATOR

### A. Target System and Virtual Platform

In the automotive industry, improvement of the electrical/electronic (E/E) architecture is a central task. The challenge therein is to find an optimal partitioning between domain-specific individual Electric Control Units (ECUs) and cross-domain ECUs [12]. At the same time, the range of functions that make use of AI algorithms is constantly growing. The wide variety of algorithmic workloads as well as a large number of different potential hardware targets raise the need for early compatibility checks and performance estimation. A versatile virtual platform, featuring easily configurable and exchangeable peripheral models facilitates the agile automotive development for such applications.

As shown in Figure 2, the VP for this case study builds on an ARM microcontroller core and peripheral models from QEMU integrated through QBox. The accelerator model is an application and workload specific design to speed-up AI workload computation by independently calculating parts of such an algorithm. Other standard peripherals are taken from the VCML library.

In this particular case, the ARM core is mainly responsible for configuring the accelerator peripheral, rendering the rather high level of abstraction of QEMU inconsequential. The accelerator processes data chunks independently, so the timing accuracy of the VP for the observed workload mainly depends on the accuracy of the peripheral, the interconnect and the memory models.

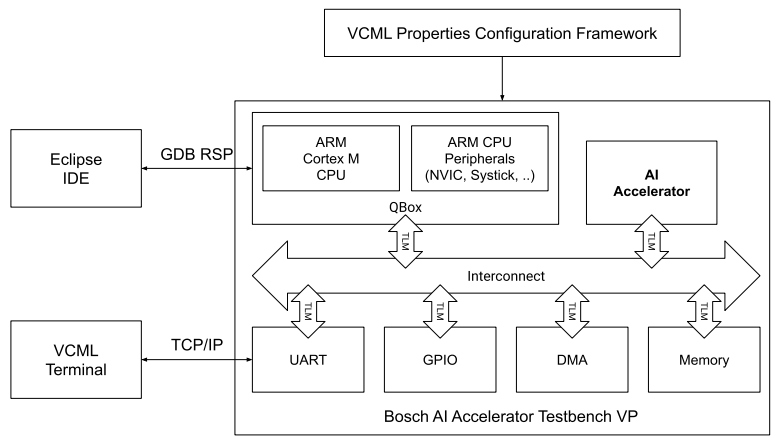

Figure 2: Block Diagram of the Virtual Prototype

### B. Software Development Flow

In the early phases of ECU development, many aspects of the overall hardware and software concept are suspect to change. Therefore, the work environment must provide intuitive configurability and enable fast test iterations. In this case study, the virtual prototype system is executed with the SUNRISE framework [11], which propagates the VCML configuration properties to the software developer through its *System API*, an interface that is agnostic to simulation technology. For the VCML/QBox VP, this encompasses hardware configuration details such as QEMU core options, custom SystemC model parameters, interconnect timing, and clock domain settings.

The compilation of the embedded software binary for the application is done with the common productive toolchain for the target core. The virtual prototype in such early project phases lacks hardware function details and components models, which are later available in the production hardware, so it is required to adapt the software accordingly (e.g. on hardware abstraction and driver level) or carve out and execute a subset of functionality. This downside is accepted in exchange for the earlier availability of the VP. The download of the application into the memory model can be done in two ways, depending on the developer's demands. The first option is downloading the binary through the configuration API as file-parameter, which is the most convenient way for non-interactive execution like regression tests or automated iterative parameter optimization. The second option is attaching to the GDB server, which is part of QEMU, with a TCP connection into the simulation container. This allows the developer to do interactive software debugging on the VP with a standard IDE.

After the simulation is finished, simulation results are retrieved through the System API. Artifacts are cycle or instruction count numbers of the QEMU ARM core, file-based artifacts like custom log files created by models in the VP, VCML trace-files that document the simulation behavior and VCD signal traces from SystemC to analyze the hardware behavior over time. The console output is also captured and contains logging outputs from the simulation as well as console output from the embedded software through the VCML UART peripheral.

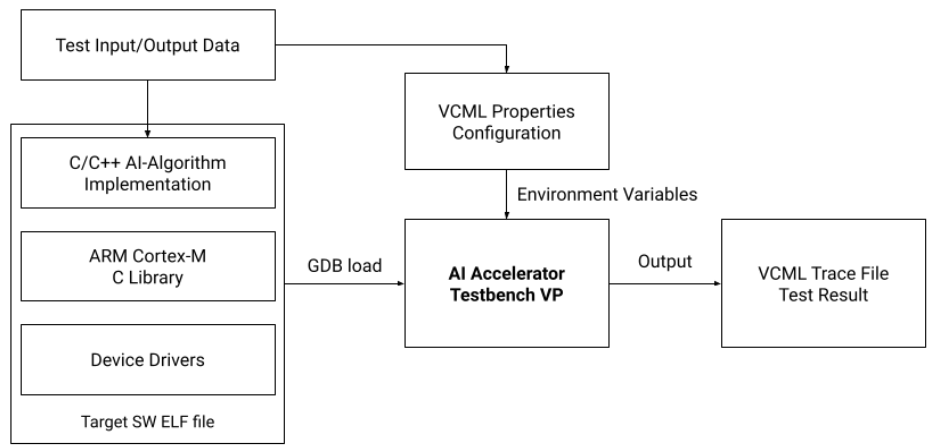

Figure 3: Target Software and Test Scenario

*C. Practical Experience*

The proposed methodology was put to test with a concrete AI workload which was executed on other virtual platforms in off-the-shelf commercial simulation tools before. This allows drawing direct conclusions about the efficiency of our approach by comparing it to the previous experience.

The virtual prototype was delivered as source code with an accompanying Dockerfile, so the simulation itself was compiled on the user-side from scratch during the build of the Docker image. Thus, the VP deliverable is very light weight and easy to handle, as it mainly contains well compressible text files instead of binaries and does not include the publicly available dependencies which can be taken from the official package repositories. This is a highly transparent approach that gives the user the opportunity to understand the build flow of the VP environment in detail, see all requirements and optionally modify parts of it. In contrast, most solutions available on the market are delivered as binary (executable) or installation package and remain a black box for the user. For the QBox-VCML VP, the only setup task was the one-time execution of the docker build command to generate the docker image which than can be re-used by all software developers, enabling instant execution of the simulation at their side with neglectable repeated effort. Overall, this means a significant improvement of the TTE. With the conventional approach in the past, setup times were in the range of days for initial setups of a new simulation technology in a project and repeated efforts in the region of half a working day for every new user. Through the containerized approach and because in the context of this work a docker engine was already set up in the development environment, the initial preparation and build took only few hours and the repeated effort is minimal. More time savings are expected in the future, if the results of the current work need to be reproduced or extended, since the docker image is archived on a local registry, so it will stay functional and can just be re-used.

An additional and underestimated source of effort for the software developer in this study was the implementation of low-level software layer that is required to execute the embedded workload on the core. In productive development for productive hardware targets, typically a board support package is already available, whereas for the custom VP a new software project with standard libraries, boot code and toolchain settings had to be created. The significant time spent on this could be counted into the TTE, since it is required to run a meaningful simulation, but it is no unique aspect for the specific technology used in this work as it is required initially for porting software on any custom system configuration. Besides the good suitability for containerized work, the QEMU model as proven to be a good choice for the use-case. The simulation performance was faster than half of real-time for the workloads tested, which are in the range of a couple of seconds simulated time. This is a convenient value of the applications targeted in this work. On the other hand, for other use-cases, that require more detailed analysis of the embedded software on the core, the QEMU model lacks timing details and tracing / profiling options, so a more capable core model on a lower abstraction level would be required.

## VI. Conclusion

By using a VP based on a QEMU CPU model in a SystemC environment with containerization technology, a highly portable and scalable simulation environment was achieved, ensuring compatibility across various platforms. Time to Execution (TTE) was significantly reduced, enhancing efficiency and minimizing set-up efforts in the development flow. The approach has been proven particularly advantageous for non-interactive, headless use cases, where automation is key. Thus, in future work the approach will be applied in continuous build and test chains for productive automotive software development.